# Giant third-order nonlinear Hall effect in misfit layer compound $(SnS)_{1.17}(NbS_2)_3$

Shengyao Li[1†], Xueyan Wang[1†*], Zherui Yang[1], Lijuan Zhang[2], Siew Lang Teo[3], Ming Lin[3], Ri He[4], Naizhou Wang[1], Peng Song[5], Wanghao Tian[5], Xian Jun Loh[3,6], Qiang Zhu[3,6,7], Bo Sun[2,8], X. Renshaw Wang[1,5*]

[1]*Division of Physics and Applied Physics, School of Physical and Mathematical Sciences, Nanyang Technological University, 21 Nanyang Link, Singapore 637371, Singapore*

[2]*Tsinghua-Berkeley Shenzhen Institute and Shenzhen Geim Graphene Center, Tsinghua University, Shenzhen, Guangdong 518055, China*

[3]*Institute of Materials Research and Engineering (IMRE), Agency for Science, Technology and Research (A*STAR), 2 Fusionopolis Way, Innovis #08-03, Singapore 138634, Singapore*

[4]*Key Laboratory of Magnetic Materials and Devices, Ningbo Institute of Materials Technology and Engineering, Chinese Academy of Sciences, Ningbo, 315201, China*

[5]*School of Electrical and Electronic Engineering, Nanyang Technological University, 50 Nanyang Ave, 639798, Singapore*

[6]*Institute of Sustainability for Chemicals, Energy and Environment (ISCE2), Agency for Science, Technology and Research (A*STAR), 1 Pesek Road, Jurong Island, Singapore 627833, Singapore*

[7]*School of Chemistry, Chemical Engineering and Biotechnology, Nanyang Technological University, 21 Nanyang Link, Singapore 637371, Singapore*

[8]*Institute of Materials Research, Tsinghua Shenzhen International Graduate School, Guangdong Provincial Key Laboratory of Thermal Management Engineering and Materials, Shenzhen, Guangdong 518055, China*

[†]These authors contributed equally
*Email: xueyan.wang@ntu.edu.sg
　　　renshaw@ntu.edu.sg

**Abstract**

Nonlinear Hall effect (NLHE) holds immense significance in recognizing the band geometry and its potential applications in current rectification. Recent discoveries have expanded the study from second-order to third-order nonlinear Hall effect (THE), which is governed by an intrinsic band geometric quantity called the Berry Connection Polarizability (BCP) tensor. Here we demonstrate a giant THE in a misfit layer compound, $(SnS)_{1.17}(NbS_2)_3$. While the THE is prohibited in individual $NbS_2$ and SnS due to the constraints imposed by the crystal symmetry and their band structures, a remarkable THE emerges when a superlattice is formed by introducing a monolayer of SnS. The angular-dependent THE and its scaling relationship indicate that the phenomenon could be correlated to the band geometry modulation, concurrently with the symmetry breaking. The resulting strength of THE is orders of magnitude higher compared to recent studies. Our work illuminates the modulation of structural and electronic geometries for novel quantum phenomena through interface engineering.

**Introduction**

Nonlinear Hall effect (NLHE) has been one of the focal topics in condensed matter physics. Unlike the necessity of broken time-reversal symmetry for the ordinary Hall effect,[1] NLHE rises from intrinsic band geometry, Berry curvature dipole (BCD).[2,3] Conventionally, NLHE manifests as a second-order transverse voltage ($V_\perp^{2\omega}$) response to a longitudinal a.c. current ($I^\omega$) in systems with broken inversion symmetry.[2] This phenomenon exhibits potential applications in recognizing the band geometry and current rectification.[4–6] Recent works extended the study to third-order nonlinear Hall effect (THE) in nonmagnetic systems with preserved inversion symmetry, where both linear and second-order Hall responses are suppressed.[7–10] The origination is intricately connected with another band geometry quantity, known as Berry connection polarizability (BCP).[7,9] Under an in-plane current excitation, BCP is responsible for an electric field-driven BCD, thereby contributing to the emergence of THE signal.[11] Being induced by the band geometry, the magnitude of BCD is significantly enhanced in tilted band anticrossings and band inversions.[3] Hence the phenomenon was experimentally reported in Weyl semimetals, such as MoTe$_2$, WTe$_2$, and TaIrTe$_4$.[7,12,13] The broad accessibility of THE position it as a potential alternative not only for fundamental band geometry studies, but also for practical applications in the detection and energy harvesting of high-frequency microwaves.[14]

However, current studies are confined to individual materials, where THE presents due to their inherent electronic and structural properties, leaving behind the investigations on interfacial structural engineering in heterostructures. Interface interaction in van der Waals materials offers an effective approach for modifying band arrangements.[15–18] Taking advantage of the interface interaction, misfit layer compounds intensified the effect by the alternate stacking of transition metal dichalcogenides (TMD) and rock salt chalcogenides.[19,20] Various novel phenomena were explored within this category, including Ising superconductivity and nonreciprocal transport.[21–24] Both experimental and theoretical works highlight the potency of manipulating the spatial and electronic structures in these misfit layer compounds.

In this work, we demonstrate the emergence of a giant THE within a misfit layer compound, (SnS)$_{1.17}$(NbS$_2$)$_3$, which is abbreviated as AB$_3$. THE is individually suppressed in both NbS$_2$ and SnS due to the constraints posed by $C_3$ symmetry and their respective band structures. Intriguingly, upon the synthetic combination of NbS$_2$ and SnS into a misfit layer compound, we observe a remarkable THE. We then attribute this phenomenon to the band geometry modulation within the misfit layer compound, where the existence of BCP facilitates the emergence of BCD under the excitation of an electric field, in concurrence with the breaking of $C_3$ symmetry.

**Results**

Band geometry plays a vital role in determining charge transport behaviours.[25] Being an intrinsic band property, BCP is a second-rank tensor, which represents the ratio between the field-induced Berry connection ($\mathcal{A}^{(1)}$) and the electric field ($E$),[10]

$$G_{ab}(k) = \frac{\partial \mathcal{A}^{(1)}}{\partial E_b}, \tag{1}$$

where $a$, $b$ denotes the Cartesian coordinates. BCP is gauge invariant, which reflects the intrinsic band properties. A typical two-dimensional (2D) gapped Dirac model (Figure 1a) is constantly employed to illustrate the origin of THE from BCP under the electric field excitation. The upper panel in Figure 1b(i) and (ii) display the distribution of BCP tensor. Similar to Berry curvature, BCP is concentrated and maximized at the small-gapped region, with monopole and quadrupole distributions for diagonal and off-diagonal components.[9] For the typical Dirac model, the intrinsic Berry curvature exhibits a monopole-like structure, concentrated at the small-gap region around the centre as shown in Figure 1b(iii). Upon the application of an electric field, the BCP contributes to a field-induced Berry connection $\mathcal{A}^E$,[9,10]

$$\mathcal{A}^E = \overleftrightarrow{G} E, \tag{2}$$

and subsequently generated a field-induced Berry curvature,

$$\Omega^E = \nabla_k \times (\overleftrightarrow{G} E). \tag{3}$$

Figure 1b(iv) illustrates the distribution of this electric field-induced Berry curvature. In contrast to the pristine monopole-like structure, this field-induced Berry curvature exhibits a dipole, with its dipole direction being perpendicular to the electric field. This field-induced Berry curvature originated from BCP, which has a natural contribution to THE.[9,14] To perform the high-order NLHE measurements, Figure 1c schematically depicts the measurement setup. $I^\omega$ is applied between the source (S) and drain (D) electrodes. High-order transverse voltages ($V_\perp^{n\omega}$) from first to third-harmonic frequencies are recorded between another pair of opposite electrodes (labelled in H and L).

Despite the widespread observation of THE in type-II Weyl semimetals due to their distinctive band structures,[7,12,13] the influence of interface interaction remains elusive. To address the issue, we investigate THE in a misfit layer compound, $AB_3$. Misfit layer compounds have been an attractive platform for realizing novel physical phenomena due to their high anisotropy and modified band structure through interlayer bonding.[21,22,26] Figure 1d displays the individual constitutes of the misfit layer compound, namely, rhombohedral phase $NbS_2$ and orthogonal phase SnS.[27] Figure 1e shows the lattice structure of $AB_3$ from the *c*-axis, the alternate stack of SnS and $NbS_2$ forms a square mesh in the *ab*-plane. These two sublayers are commensurate along *b* and *c* crystallographic directions, but incommensurate along *a* crystallographic direction, which determined the composition.[28] By applying a low-frequency $I^\omega$ with a direction $\theta$ degree away from the crystal *b* axis, we investigate the anisotropy of the superlattice. Figure 1f shows the high-resolution transmission electron microscopy (HRTEM) of $AB_3$, where the two sublayers are distinguished by different contrast. The blue shaded area corresponds to $NbS_2$, while the orange shaded area represents SnS.

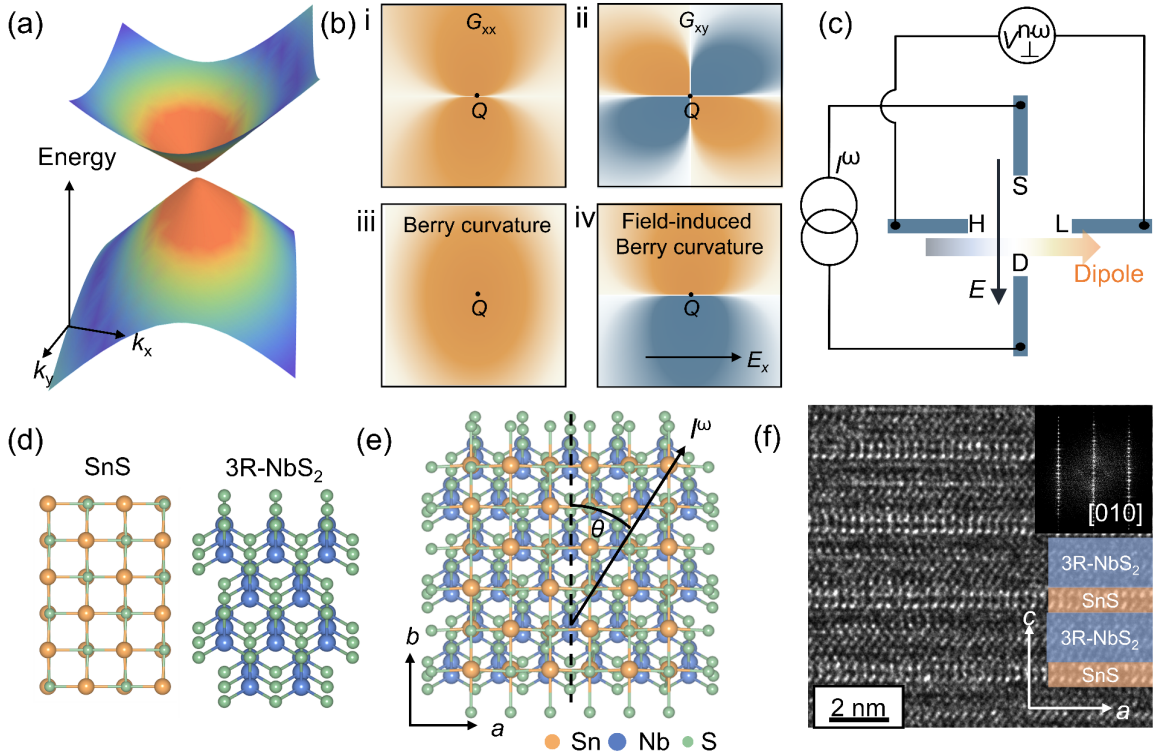

**Figure 1.** Berry connection polarizability (BCP) in a two-dimensional (2D) gapped Dirac model and crystal structure of $(SnS)_{1.17}(NbS_2)_3$ (abbreviated as $AB_3$). (a) Band structure of a 2D gapped Dirac model. (b) Distribution of BCP and Berry curvature. (i) and (ii), BCP components concentrated and maximized at the Dirac point. (iii) Monopole structure of intrinsic Berry curvature. (iv) Dipole structure of the electric field induced Berry curvature. (c) Electrode configuration for the nonlinear Hall effect (NLHE) measurement. Current is applied between the S and D electrodes; Hall responses of different orders are measured in the transverse direction between H and L electrodes. (d) *ab*-plane of a monolayer of SnS and $NbS_2$. (e) *ab*-

plane of AB$_3$. The orange, green, and blue balls stand for Sn, S and Nb, respectively. An a.c. current ($I^\omega$) is applied $\theta$ degree away from the crystal $b$ axis. (f) Cross-sectional high-resolution transmission electron microscopy (HRTEM) image along the [010] direction. The NbS$_2$ layer (dark) and SnS (bright) layer can be distinguished by different contrast. Scale bar: 2 nm.

Figure 2a displays the optical image of the device. The thickness of the flake is 14 nm, with Cr/Au electrodes patterned on top of the flake. $I^\omega$ is applied between the S and D electrodes, and $V_\perp^{n\omega}$ is measured between H and L electrodes. Figure 2b shows the THE response upon swapping the current electrodes. The sign reversal of THE response is attributed to the alteration in BCD direction when subjected to an opposite electric field. Further swapping voltage electrodes also leads to a sign reversal in THE, indicating the presence of a dipole along the transverse direction (Figure 2c). To confirm that the observed THE is a consequence of the interface interaction between NbS$_2$ and SnS, rather than an intrinsic property of an individual constitute, we provide a comparative experiment between AB$_3$ and NbS$_2$, as illustrated in Figure 2d-f. Note that we exclude SnS from this discussion due to its semiconductor characteristics.

Figure 2d shows the variation of third-order transverse voltage, $V_\perp^{3\omega}$, as a function of the longitudinal voltage, $V_\parallel^\omega$. $V_\perp^{3\omega}$ has a nonlinear dependency on $V_\parallel^\omega$ in AB$_3$ but is negligible in NbS$_2$. Figure 2e shows the first-order transverse voltage $V_\perp^\omega$ versus $V_\parallel^\omega$. In principle, Hall voltage is constrained by the preserved time-reversal symmetry. The observed linear relationship between $V_\perp^\omega$ and $V_\parallel^\omega$ is the resistance contribution from the misalignment of electrodes and the intrinsic anisotropic resistance from AB$_3$ crystal. Figure 2f shows the variation of $V_\perp^{2\omega}$ versus $V_\parallel^\omega$. In AB$_3$, $V_\perp^{2\omega}$ is an order smaller than the magnitude $V_\perp^{3\omega}$, indicating a dominant contribution from the electric field-induced BCD, rather than the intrinsic BCD, that contributes to the NLHE. According to the mechanism of THE, the third-order Hall current response is suppressed by a $C_3$ symmetry in NbS$_2$.[9] The constraint of $\chi_{11} = \chi_{22} = 3\chi_{12} = 3\chi_{21}$ suppress $V_\perp^{3\omega}$ response down to zero in NbS$_2$.[9] The observation of $V_\perp^{2\omega}$ and $V_\perp^{3\omega}$ in AB$_3$ indicates that the formation of a misfit layer compound greatly changes the crystal structure and the band geometry.[28–30]

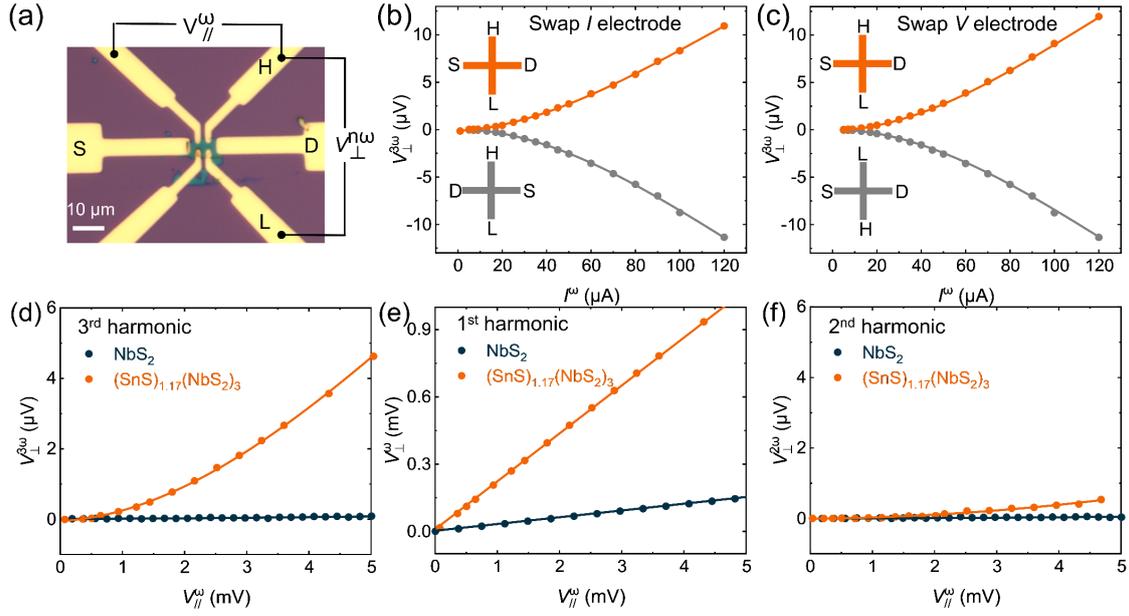

**Figure 2.** NLHE of AB$_3$ at $T$ = 2 K. (a) Optical image of a Hall bar device. Scale bar: 10 μm. $I^\omega$ is applied between S and D electrodes; Hall voltage is measured in the transverse direction between H and L electrodes. (b) third-order transverse voltage ($V_\perp^{3\omega}$) changes sign when swapping S and D electrodes, indicating electric fields of opposite directions reverse Berry curvature dipole (BCD). (c) $V_\perp^{3\omega}$ changes sign when swapping H and L electrodes, indicating the BCD direction being transverse to the electric field. (d) $V_\perp^{3\omega}$ versus first-order longitudinal voltage ($V_\parallel^\omega$) of AB$_3$ and NbS$_2$ at $T$ = 2 K. THE response is below the detectable limit in NbS$_2$. (e) First-order transverse voltage ($V_\perp^\omega$) versus $V_\parallel^\omega$ of AB$_3$ and NbS$_2$ at $T$ = 2 K. The presence of ordinary Hall response is from the electrode misalignment and crystal anisotropy. (f) Second-order transverse voltage ($V_\perp^{2\omega}$) versus $V_\parallel^\omega$ of AB$_3$ and NbS$_2$ at $T$ = 2 K. $V_\perp^{2\omega}$ is relatively small in both materials, indicating the dominant contribution from THE.

The distribution of THE is intricately correlated with the crystal symmetry. Figure 3a schematically illustrates the angle-dependent measurement. $I^\omega$ is applied between two opposite electrodes. Then, $V_\perp^{n\omega}$ is measured between a pair of electrodes perpendicular to the current direction, and $V_\parallel^\omega$ is measured between another pair of electrodes parallel to the current direction. By anticlockwise rotating the measurement framework while maintaining the relative position unchanged, we obtain the angle-dependent properties of AB$_3$. Figure 3b displays the circular disk-like device for the angle-dependent measurements. An exfoliated flake was transferred onto the pre-patterned circular disk shape electrodes, with each neighbouring electrode positioned at a 30° separation. Each electrode is aligned along the radius and the flake is cut into a star shape using focused ion beam (FIB) cutting for the angle-resolved measurements. The crystal $b$ ($a$) axis is along the long (short) side, confirmed after the measurements (Figure S5, Supporting Information). We then define $\theta$ as the $I^\omega$ direction starting from the crystal $b$ axis.

Figure 3c shows $V_\parallel^\omega$ versus $I^\omega$ of different directions. $I^\omega$ is applied between opposite electrodes along the radial direction, and $V_\parallel^\omega$ is measured at the adjacent electrodes. The linear relationship between $V_\parallel^\omega$ and $I^\omega$ indicates a good contact between the gold electrode and the material. And the periodic variation in slopes indicates an anisotropic resistance arising from the crystal structure. To determine the anisotropic resistance ($R_\parallel^\omega$) of AB3, Figure 3d plots $R_\parallel^\omega$ as a function of $\theta$. $R_\parallel^\omega$ exhibits two-fold symmetry, where the resistance is maximized along the crystal $b$ axis and minimized along the crystal $a$ axis. The curve can be fitted by

$$R_\parallel^\omega(\theta) = R_b \cos^2\theta + R_a \sin^2\theta, \qquad (4)$$

where $R_a$ and $R_b$ are the resistance along the crystal $a$ and $b$ axes ($R_b > R_a$), respectively. The resistance anisotropies of the crystal, $r \equiv R_a/R_b = 0.76$. $V_\perp^{n\omega}$ is measured in the direction perpendicular to the current. Figure 3e displays $V_\perp^\omega$ versus current of different directions, and the angle-dependent Hall resistance ($R_\perp^\omega$) is shown in Figure 3f. The data points can be well-fitted by

$$R_\perp^\omega(\theta) = (R_b - R_a)\sin\theta\cos\theta. \qquad (5)$$

$V_\perp^\omega$ increases linearly with $I^\omega$ for the electric field of all the directions, which is from the slight electrode misalignment and the resistance anisotropy. Upon confirming the highly anisotropic structure of AB3, we then focus on the angle dependence of THE. Figure 3g shows that $V_\perp^{3\omega}$ scales linearly with $(V_\parallel^\omega)^3$ for all the directions. The slope of $V_\perp^{3\omega}$ versus $(V_\parallel^\omega)^3$ as a function of $\theta$ is shown in Figure 3h. According to principle,[9] this THE originated from BCP is perpendicular to the electric field, which keeps sign and amplitude by simultaneously reversing the direction of the electric field and voltage probes. Under the constraint of mirror symmetry, the $V_\perp^{3\omega}$ response vanishes when the electric field is along the high-symmetry axis. The data points can then be fitted by

$$\frac{V_\perp^{3\omega}}{(V_\parallel^\omega)^3} \propto \frac{\cos\theta\sin\theta[(3\chi_{21}r^2 - \chi_{11})\cos^2\theta + (\chi_{22}r^4 - 3\chi_{12}r^2)\sin^2\theta]}{(\cos^2\theta + r\sin^2\theta)^3}, \qquad (6)$$

where $\chi_{ij}$ are the elements of the third-order susceptibility tensor, $r$ is the resistance anisotropy.

This anomalous angular dependence of the $V_\perp^{3\omega}/(V_\parallel^\omega)^3$ can be explained by the electric field-induced BCD. The angle-dependent THE also demonstrates a two-fold symmetry. The theory and the fitting curve are consistent with our experimental data, where THE vanishes when the driving field is parallel with the crystal high symmetric $a$ and $b$ axis of AB$_3$.

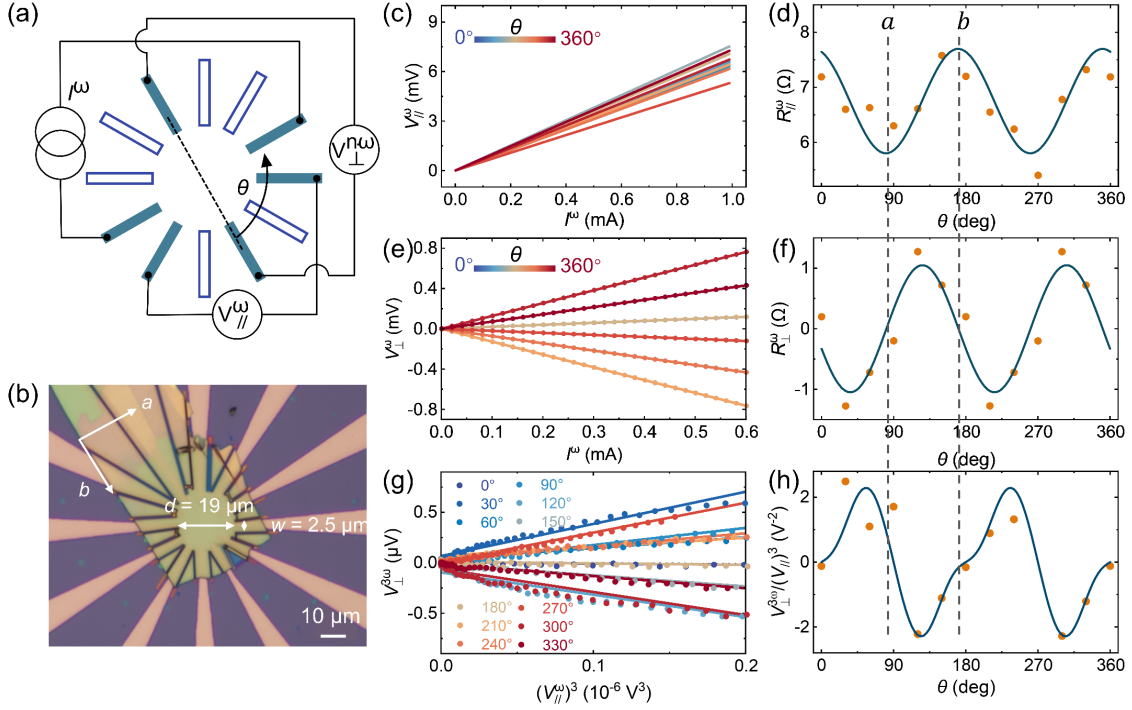

**Figure 3.** Angle-dependent NLHE of AB$_3$ at $T$ = 2 K. (a) Schematic illustration of the angle-dependent measurement. $V_\parallel^\omega$ and $V_\perp^{3\omega}$ are measured simultaneously. (b) The optical image of a device with 12 electrodes spaced 30° apart. Scale bar: 10 μm. The diameter of the circular disk, $d$ = 19 μm, and the width of each finger, $w$ = 2.5 μm. (c) $V_\parallel^\omega$ versus $I^\omega$ of different directions, with $\theta$ degree away from crystal $b$ axis. (d) Anisotropic resistance ($R_\parallel^\omega$) as a function of $\theta$. (e) $V_\perp^\omega$ versus $I^\omega$ of different directions. (f) Hall resistance ($R_\perp^\omega$) as a function of $\theta$. (g) $V_\perp^{3\omega}$ versus $(V_\parallel^\omega)^3$ of different directions, fitted by linear lines. (h) Variation of $V_\perp^{3\omega}/(V_\parallel^\omega)^3$ as a function of $\theta$, showing an angular dependence on the crystal structure.

The emergence of THE comprises various contributions, which include the electric field-induced BCD and scattering.[12] Given the metallic property of AB$_3$, we control carrier scattering by adjusting the temperature. Figure 4a presents that $V_\perp^{3\omega}$ scales linearly with $(V_\parallel^\omega)^3$ for all the temperatures. The absolute value of the slope for each curve, $V_\perp^{3\omega}/(V_\parallel^\omega)^3$ decreases monotonically with increasing temperature, finally being suppressed to nearly zero above 25 K

(Figure 4b). To elucidate the scaling relation in terms of intensive quantities, the voltage is transformed to an electric field using $E_\perp^{3\omega} = V_\perp^{3\omega}/d$. The conductivity $\sigma$ is calculated as $\sigma = \frac{1}{\rho} = \frac{d}{A*R_\parallel^\omega}$, where A is the cross-sectional area. Figure 4c displays the variation of $\frac{E_\perp^{3\omega}}{(E_\parallel)^3}$ as a function of $\sigma^2$, which demonstrate a scaling relationship (grey dashed line) as

$$\frac{E_\perp^{3\omega}}{(E_\parallel)^3} = \xi\sigma^2 + \eta. \tag{6}$$

Note that $\frac{E_\perp^{3\omega}}{(E_\parallel)^3} \propto \frac{\chi}{r\sigma}$, where $\tau$ is the scattering time, $\chi$ is the third-order Hall susceptibility, $r$ is the resistance anisotropy. Considering the linear dependency of $\tau$ on $\sigma$, we obtain that $\chi$ has two contributions, which scales linearly with $\tau$ and $\tau^3$. In conventional magnetic materials, the Hall conductivity for intrinsic Berry curvature and side-jump contributions scales as $\tau^0$,[31] while the skew-scattering terms scale as $\tau^2$.[32] Combining with the linear dependency of $\tau$ on the external electric field, these factors jointly contributes to the scales in $\tau$ and $\tau^3$.

In the perspective of $\tau$, the intrinsic THE contribution from the field-induced polarization scales as $\tau$, which consists with the linear dependency of $\tau$ in our scaling relationship. Though the extrinsic contribution to THE from side-jump scattering also scales with $\tau$, the high mobility and low resistivity indicate a rather weak contribution from side-jump scattering.[31] Moreover, the angle dependence of THE is well-fitted with the crystal symmetry of AB$_3$, which is not likely to originate from extrinsic impurities. Other extrinsic contributions to THE, such as the capacitance coupling and thermal effect, are discussed in Supporting Information. In the regards of $\tau^3$, both the skew-scattering and Drude-model could be responsible for the linear dependency. As illustrated in recent works,[12,13] considering the polarization of Bloch electrons with orbital magnetic moment, the opposite polarized Bloch electrons will experience an asymmetric scattering,[33] contributing to a transverse current.

Compared with recent works, the THE is orders of magnitude higher even in bulk crystal of AB$_3$ (Figure 4d). This enhancement may arise from distinctive features of misfit layer compounds. The spatial structure is changed through the alternate stacking of sublayers, thereby influencing the electrical and optical transport properties. In AB$_3$, though THE is suppressed in individual NbS$_2$ due to $C_3$ symmetry, the intercalation of SnS sublayer modifies the crystal structure to a two-fold symmetry, thereby giving rise to THE. Moreover, the alternate stacking of sublayers functions as a periodic field effect transistor, with charges transfer between sublayers beyond the modulation effect of an external electric field.[29,30] The unique characteristics of misfit layer compounds also give rises to other intriguing physical phenomena such as Ising superconductivity, but still posing questions about the underlying mechanism.[21] Hence the observation of the giant THE in $AB_3$ provides valuable insights into the electronic properties of misfit layer compounds.

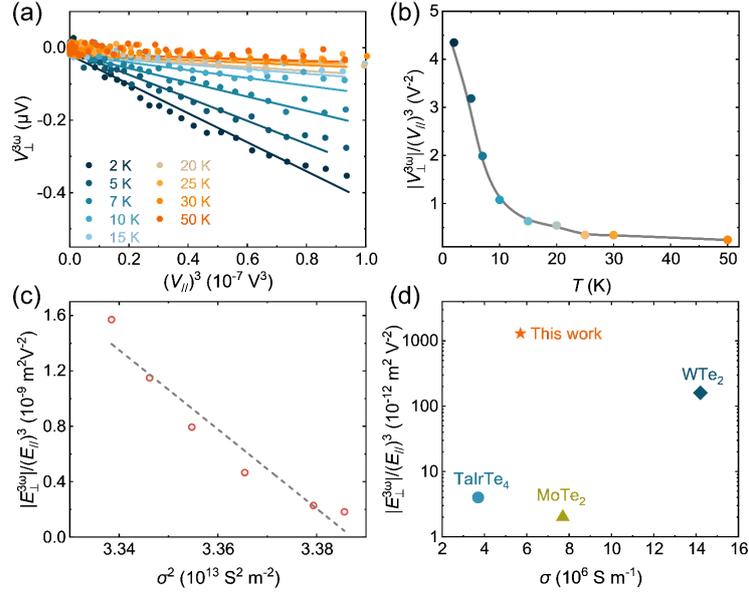

**Figure 4.** Temperature dependence of THE. (a) THE at different temperatures ranging from 2 to 50 K. (b) Temperature dependence of $V_\perp^{3\omega}/(V_\parallel^\omega)^3$. The amplitude exhibits an exponential decrease, approaching nearly zero at temperatures above 25 K. (c) Absolute value of $E_\perp^{3\omega}/(E_\parallel)^3$ versus the square of longitudinal conductivity ($\sigma^2$). The scaling relationship indicates that THE has contributions from two terms. (d) Giant THE compared with recent works in Weyl semimetals.

**Conclusion**

In summary, we demonstrate a giant THE originated from the interfacial interaction within a misfit layer compound. In contrast to the lack of THE in $NbS_2$ due to the constraint on the crystal and electronic structure, the misfit layer compound, $(SnS)_{1.17}(NbS_2)_3$, presents a giant THE by an order higher compared with recent works. Through investigations into the angle dependence of THE and its scaling relationship, we found that the BCP-induced BCD under an in-plane electric field is responsible for this THE. Therefore, the alternate stack of SnS and $NbS_2$ not only breaks the crystal symmetry but also could modulate the band geometry, where

BCP is enhanced. This work experimentally enlightens the acquisition of polarization in momentum space by tailoring the structural and electronic geometry in a misfit layer compound.

## Methods

*Crystal synthesis and characterization*

$(SnS)_{1.17}(NbS_2)_3$ crystals were synthesized using an iodine vapour transfer method. A mixture of stoichiometric elementary substances was loaded into a quartz tube within an argon-filled glove box. Subsequently, the tube underwent evacuation, flame-sealing, and sintering in a tube with a temperature gradient from 770 to 850 °C. After four weeks, the sintering yields plate-like crystals up to 5 mm.

The crystal structures were characterized by XRD (Bruker D8 Advance) using Cu Kα radiation with λ = 1.5418 Å and a LynxEye detector. The XRD analysis was conducted in the conventional Bragg-Brentano configuration, spanning a 2θ range from 10° to 80°, with a step size of 0.02°. HRTEM was performed on the JEOL JEM 3200FS at 300 kV. Preparation of thin TEM specimens involved the use of a focused ion beam (FIB) system at a moderate voltage (30 keV). Subsequent cleaning steps were carried out using low voltages of 5 and 2 keV.

*Device fabrication*

For the preparation of the disk-like device, the $(SnS)_{1.17}(NbS_2)_3$ flakes were mechanically exfoliated from bulk crystals onto the oxygen plasma-treated $SiO_2$/Si substrates. The flakes were dry transferred using polycarbonate (PC) stamps onto the pre-patterned Cr (5 nm) /Au (20 nm) bottom electrodes. Then, the device was fabricated into a star shape using the focused ion beam (FIB) cutting. For the preparation of the Hall bar device, due to the hard-to-exfoliation characteristic, Au-assisted exfoliation was applied to get thin flakes. Au (100 nm) was first deposited on $SiO_2$/Si substrates using a thermal evaporator. After cutting into small pieces, the small pieces were stuck to the glass slides using the UV glue with face down. After solidifying for 4 hrs under a UV light of 30 W, the $SiO_2$/Si substrate was removed, leaving an atomic flat Au film on glass slides. Large-size thin flakes were obtained on this atomic flat Au film. The thin flakes were then transferred to $SiO_2$/Si substrate, patterned into a Hall bar structure using UV lithography via an Ultraviolet Maskless Lithography machine (TuoTuo Technology

(Suzhou) Co., Ltd.). Cr (5 nm)/ Au (50 nm) top electrodes were then deposited using a thermal evaporator.

*Electrical measurements*

The low-temperature transport measurements were performed in an Oxford TeslatronPT cryostat. The sample was loaded on a rotation probe from MultiField for the electrical measurements. The temperature-dependent resistance curves were measured using a Keithley 6221 triggered with a Keithley 2182, at a frequency of 21 Hz. In the measurement of high-order nonlinear Hall effect, the a.c. current was applied using a Keithley 6221 at a frequency of 137.7 Hz, and the SR830 lock-in amplifier was used to monitor the voltage drop. During the sweeping of the current amplitude, five data points were obtained at each current amplitude and took the average.

**Acknowledgement**

L. Zhang acknowledges the support from the National Natural Science Foundation of China (No. 52002206). X.R.W. acknowledges support from Singapore Ministry of Education under its Academic Research Fund (AcRF) Tier 1 (Grant No. RG82/23), Tier 2 (Grant Nos. MOE-T2EP50120-0006 and MOE-T2EP50220-0005) and Tier 3 (Grant No. MOE2018-T3-1-002), and the Agency for Science, Technology and Research (A*STAR) under its AME IRG grant (Project No. A20E5c0094). X.R.W. designed and directed this study. We acknowledge the discussion and support from Prof. Weibo Gao.

**Supporting Information**

The Supporting Information contains the detailed structure of $(SnS)_{1.17}(NbS_2)_3$, the verification of the intrinsic contribution of THE, THE responses in other samples without FIB cutting and the impact of electric field on the second-order nonlinear Hall effect.

Table of Contents

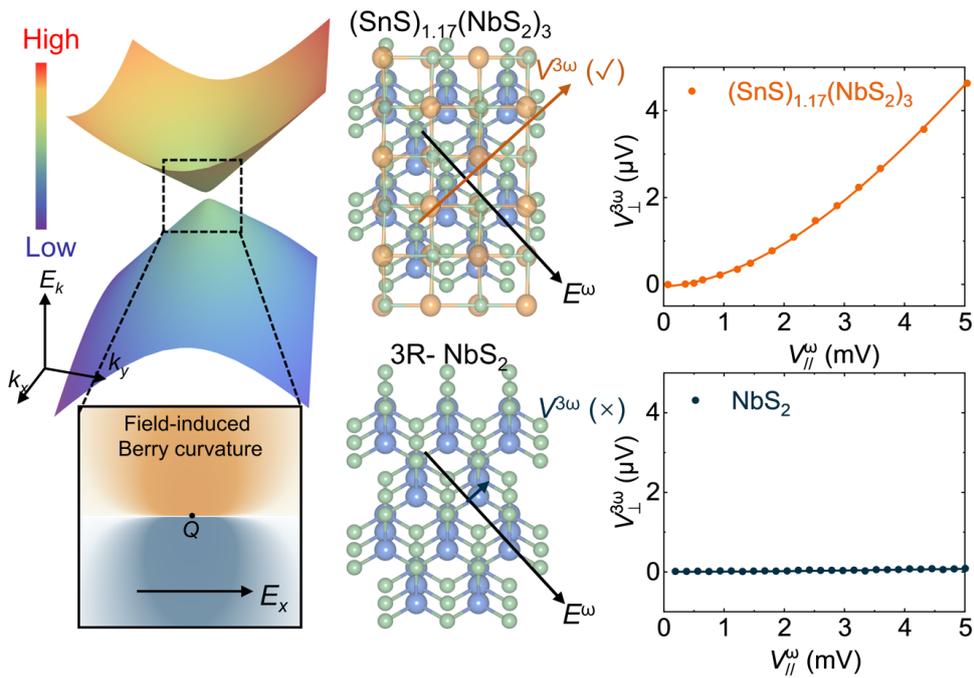